\documentclass[12pt,a4paper,fleqn]{article}

\usepackage[DIV12]{typearea}
\usepackage{amsmath,amsfonts,amssymb}
\usepackage{graphicx}
\usepackage[footnotesize]{caption2}
\usepackage{cite}

\newcommand{\stau}{{\widetilde\tau}}
\newcommand{\snu}{{\widetilde\nu}}
\newcommand{\slepton}{{\widetilde\ell}}
\newcommand{\gravitino}{\widetilde{G}} 

\newcommand{\Eqref}[1]{Eq.~\eqref{#1}}
\newcommand{\Figref}[1]{Fig.~\ref{#1}}
\newcommand{\Tabref}[1]{Tab.~\ref{#1}}

\newcommand{\eVdist}{\kern-0.06em}

\newcommand{\Gev}{\text{Ge\eVdist V}}
\newcommand{\Tev}{\text{Te\eVdist V}} 

\newcommand{\gev}{\:\text{Ge\eVdist V}}
\newcommand{\tev}{\:\text{Te\eVdist V}}
\newcommand{\s}{\:\text{s}}

\allowdisplaybreaks[1]

\begin{document}

\begin{titlepage}

\begin{flushright}
TUM-HEP 679/07
\end{flushright}

\vspace*{1.0cm}

\begin{center}
{\Large\bf 
The Gravitino-Stau Scenario after Catalyzed BBN
}

\vspace{1cm}

\textbf{
J\"orn Kersten\footnote[1]{Email: \texttt{jkersten@ictp.it}}$^{,\,a}$
and
Kai Schmidt-Hoberg\footnote[2]{Email: \texttt{kai.schmidt-hoberg@ph.tum.de}}$^{,\,b}$
}
\\[5mm]
\textit{\small
$^{a}$%
The Abdus Salam ICTP, Strada Costiera 11, 34014 Trieste, Italy\\[2mm]
$^{b}$%
Physik-Department T30, Technische Universit\"at M\"unchen, \\
James-Franck-Stra\ss e, 85748 Garching, Germany
}
\end{center}

\vspace{1cm}

\begin{abstract}
\noindent 
We consider the impact of Catalyzed Big Bang Nucleosynthesis on theories
with a gravitino LSP and a charged slepton NLSP\@.  In models where the
gravitino to gaugino mass ratio is bounded from below, such as
gaugino-mediated SUSY breaking, we derive a lower
bound on the gaugino mass parameter $m_{1/2}$.  As a concrete example,
we determine the parameter space of gaugino mediation that is compatible
with all cosmological constraints.
\end{abstract}

\end{titlepage}

\newpage

\section{Introduction}

The observed primordial abundances of light elements produced in Big
Bang Nucleosynthesis (BBN) allow to place stringent constraints on 
supergravity theories with conserved R parity.  Due to the extremely
weak coupling of the gravitino, there is likely a
long-lived particle whose decays happen during or after BBN and induce
nuclear reactions that change the element abundances
\cite{Falomkin:1984eu,Khlopov:1984pf,Ellis:1984eq}.  If this particle
is the gravitino itself, which is the case in the standard scenario with
a neutralino LSP, either it has to be very heavy or the
reheating temperature has to be rather low \cite{Kawasaki:2004qu}.  An attractive alternative
is to make the gravitino the LSP\@.  Then BBN is endangered by late decays
of the next-to-lightest superparticle (NLSP)\@.  
This yields significant constraints for gravitino masses in the
\Gev{} range, which are expected in gravity and gaugino mediation, for
example.
A neutralino NLSP is excluded \cite{Feng:2004mt,Cerdeno:2005eu}.
Scenarios with a sneutrino NLSP are essentially unconstrained
but very hard to test experimentally \cite{Kanzaki:2006hm,Covi:2007xj}.
Therefore, a charged slepton is a particularly interesting NLSP
candidate, the more so as this might allow
for an indirect observation of the gravitino at colliders
\cite{Buchmuller:2004rq} and neutrino telescopes
\cite{Albuquerque:2003mi}.
The slepton NLSP abundance and lifetime can satisfy the limits obtained
from BBN by considering NLSP decays alone
\cite{Fujii:2003nr,Feng:2004mt,Ellis:2003dn,Roszkowski:2004jd,Cerdeno:2005eu,Jedamzik:2005dh,Steffen:2006hw,Buchmuller:2006nx,Cyburt:2006uv,Pradler:2006hh}.
However, it was recently discovered that there is another process
involving long-lived charged particles, which was called Catalyzed BBN
(CBBN)\@.  Charged NLSPs form bound states with light nuclei, which
leads to a drastic change of some reaction rates resulting in an
overproduction of $^6$Li \cite{Pospelov:2006sc}.
A number of works 
\cite{Pospelov:2006sc,Kohri:2006cn,Kaplinghat:2006qr,Cyburt:2006uv,%
Hamaguchi:2007mp,Bird:2007ge,Kawasaki:2007xb,Jedamzik:2007cp,Pradler:2007is}
have studied this effect, typically finding upper bounds of a few
thousand seconds on the NLSP lifetime, unless its relic abundance is a
lot smaller than what is generically expected with supersymmetry
and the standard cosmology.

In order to obtain such a short lifetime, a relatively heavy
superpartner mass spectrum with a large hierarchy between NLSP and
gravitino mass is required.  In constrained scenarios for SUSY breaking
like the CMSSM, this leads to a lower bound on the gaugino mass
parameter $m_{1/2}$, which depends on the gravitino mass as long as the
latter is a free parameter \cite{Pradler:2007is} (see also
\cite{Cyburt:2006uv,Pradler:2006hh}).  Thus, the bound can in principle
be avoided by lowering $m_{3/2}$ sufficiently.  In the following, we
study the impact of CBBN constraints on the slepton NLSP region of
SUSY-breaking scenarios where the ratio of gravitino and gaugino mass is
bounded from below and where the gaugino masses unify at the GUT scale.
In this case, there is an absolute lower bound on $m_{1/2}$.  As a
specific example, we consider gaugino mediation
\cite{Kaplan:1999ac,Chacko:1999mi}.  The situation should be similar in
concrete models for gravity mediation that establish a relation between
$m_{3/2}$ and other mass parameters such as $m_{1/2}$ or the universal
scalar mass $m_0$.  We determine the parameter space of gaugino
mediation for moderate values of $\tan\beta$ that leads to a charged
slepton NLSP and is allowed by all cosmological constraints, i.e.\ the
bound on the NLSP lifetime, the bound on the energy release in decays,
and the observed dark matter density.

We will start by reviewing the parameter space of gaugino mediation in
the next section.  Afterwards, we will consider consequences of the
bound from CBBN on the NLSP lifetime, first in a more general setup and
then applied to gaugino mediation.  Adding the other cosmological
constraints, we will numerically determine the parameter space that
remains allowed and briefly discuss phenomenological consequences.

\section{Gaugino Mediation and its Parameter Space}

The scenario of gaugino-mediated SUSY breaking \cite{Kaplan:1999ac,Chacko:1999mi}
postulates the existence of $D-4$ extra spatial dimensions, which are
compactified with radii $\sim 1/M_C$.  
At different positions in the compact dimensions,
four-dimensional branes are located.
The gauge 
superfields and the Higgs fields live in the bulk.
The superfield responsible for SUSY breaking
is localised on one of the branes, while the remaining
MSSM fields are localised on different branes.  As a consequence, only
the bulk fields obtain SUSY-breaking soft masses at the compactification
scale $M_C$.  Assuming gauge coupling unification and 
$M_C \sim M_\text{GUT}$, one thus obtains the boundary conditions
\begin{subequations} \label{eq:gauginomed}
\begin{align}
&g_1 = g_2 = g_3 = g \simeq 1/\sqrt{2} \;,
\\
  &M_1 = M_2 = M_3 = m_{1/2} \;,
\\
  &m_{3/2} \neq 0 \;,
\\
  &\mu, B\mu, m^2_{h_i} \neq 0 \quad (i=1,2) \;,
\\
  &m_{\tilde \phi_\text{L}}^2 =
  m_{\tilde \phi_\text{R}}^2 = A_{\tilde\phi} = 0
  \quad \text{for all squarks and sleptons } \tilde\phi \;,
\end{align}
\end{subequations}
where the GUT charge normalisation is used for $g_1$ and where $h_1$ is
the Higgs which couples to the down-type quarks, whereas $h_2$ is the
up-type Higgs.  We have neglected small corrections from gaugino loops
and brane-localised terms breaking the unified gauge symmetry.
Eqs.~\eqref{eq:gauginomed} are valid at the compactification scale.  The
renormalisation group running to low energies generates non-zero squark
and slepton masses as well as $A$ terms, so that a realistic mass
spectrum can be obtained.

Taking the boundary conditions \eqref{eq:gauginomed} at the GUT scale,
the resulting allowed parameter space leads to several different
candidates for the (N)LSP \cite{Buchmuller:2005ma,Evans:2006sj}.
Besides the lightest neutralino, the lightest MSSM superparticle can be
a stau, a selectron or a sneutrino.  As the latter particles are not
viable dark matter candidates, they can only be the NLSP, with the
gravitino as the LSP, as long as R parity is conserved.  We will be
interested in charged NLSPs only.  The corresponding parameter space
region, usually denoted as charged slepton or $\slepton$ region in the
following, lies around the origin in the plane of the soft Higgs masses
$m_{h_1}^2$ and $m_{h_2}^2$.
Below its lower end (for too small $m_{h_1}^2$), there are no physical
points.  On the other sides, the region is bounded by the neutralino
LSP domain.  The charged sleptons are predominantly composed of the
superpartners of the right-handed leptons here.  The squark and slepton
mass spectrum depends on the soft Higgs masses only via the combination
$m^2_{h_1}-m^2_{h_2}$ to a good approximation.  The effect of a change
of $m_{1/2}$ is mainly a rescaling of the mass spectrum and of the
charged slepton region.
For moderate values of $m^2_{h_2}$, the lightest
neutralino is bino-like.  For large positive $m^2_{h_2}$, the $\mu$
parameter becomes small, so that there is a sizable higgsino admixture
in the lightest neutralino.  A selectron is the NLSP in the lower part
of the $\slepton$ region, where at least one soft Higgs mass is
negative.  For larger values of $\tan\beta$, there is also a parameter
space region where a predominantly ``left-handed'' charged slepton is
the NLSP, but we will not study this option in detail here.

The $\slepton$ region includes areas with negative $m_{h_1}^2$ or
$m_{h_2}^2$.  In parts of these areas, the scalar
potential may have charge and colour breaking minima, and the GUT
stability constraint $\mu^2(M_\text{GUT})+m_{h_{1,2}}^2(M_\text{GUT})>0$
that is invoked to avoid electroweak symmetry breaking at high energies
may be violated \cite{Ellis:2002iu,Baer:2004fu,Evans:2006sj}.  To be
conservative, we do not impose such constraints (see e.g.\
\cite{Falk:1995cq,Riotto:1995am,Kusenko:1996jn,Falk:1996zt} for
discussions of their applicability).
They would mainly affect the selectron NLSP region.

The gravitino cannot be arbitrarily light.  Using na\"ive dimensional
analysis \cite{Chacko:1999hg}, one can estimate $m_{3/2} \gtrsim 0.1 \,
m_{1/2}$ for $D=6$, $M_C \sim 2 \cdot 10^{16}\gev$ and a cutoff for the
extra-dimensional theory at the $D$-dimensional Planck scale
\cite{Buchmuller:2005rt}.  As NDA yields only a rough estimate of the
lower limit, it does not appear unreasonable to violate it by say
up to an order of magnitude.  The bound can also be relaxed by
increasing the number of compact dimensions%
\footnote{Note that the bound was derived for compact dimensions of
equal size, which may be disfavoured for larger $D$
\cite{Hebecker:2004ce}.
}
or by lowering the cutoff of the $D$-dimensional theory.  On the other
hand, $D=5$ or $M_C<M_\text{GUT}$ yields a larger gravitino mass.  For
example, changing only $D$ to $5$ yields 
$m_{3/2} \gtrsim 0.2 \, m_{1/2}$.  The case $M_C>M_\text{GUT}$ is less
interesting in our context, since then the running above the unification
scale tends to make the stau heavier than the
lightest neutralino \cite{Schmaltz:2000gy,Schmaltz:2000ei}.

\section{Constraints from Catalyzed BBN}
\subsection{Estimate of the Minimal Gaugino Mass}

No cosmological constraints have been taken into account so far.  As
mentioned, CBBN places very stringent bounds on scenarios with
long-lived charged particles \cite{Pospelov:2006sc}.  We assume the
standard cosmological scenario where the NLSP abundance equals its
thermal relic abundance, determined at the time when the particle
decouples from thermal equilibrium.  In particular, we assume it to be
in thermal equilibrium at early times and no significant entropy
production after decoupling that would dilute the abundance.  Then, the
abundance in supersymmetric theories generically exceeds the bound from
CBBN by orders of magnitude, if the NLSP lifetime is larger than
$10^3$\,--\,$10^4\s$.  Consequently, the only possibility is to decrease
the lifetime to values where the NLSP decays before the catalysis can be
completed.  As the corresponding upper limit on the lifetime is still
somewhat uncertain
\cite{Pospelov:2006sc,Cyburt:2006uv,Hamaguchi:2007mp,Bird:2007ge,%
Kawasaki:2007xb,Jedamzik:2007cp,Pradler:2007is}, 
we use a conservative value of
\begin{equation} \label{eq:CBBNBound}
	\tau_\slepton \lesssim \tau_\slepton^\text{max} = 5 \cdot 10^3\s \;.
\end{equation}

The decay rate of a charged slepton NLSP is dominated by the two-body
decay into lepton and gravitino,
\begin{equation}\label{eq:2bodydecay}
 \Gamma_{\slepton} =
 \frac{m_{\slepton}^5}{48\pi\,m_{3/2}^2\,M_\text{P}^2} \,
	\biggl(1-\frac{m_{3/2}^2}{m_{\slepton}^2}\biggr)^4 \;,
\end{equation}
where $m_{\tilde{l}}$ is
the slepton mass, $M_\text{P} = 2.44 \cdot 10^{18} \gev$ is the reduced
Planck mass, and where the lepton mass has been neglected.
In order to minimise the lifetime, we have to
\begin{enumerate}
\item maximise the NLSP mass, i.e.\ it should be just
 below the mass of the next-heavier particle, the lightest neutralino,
 and to
\item minimise the gravitino mass.
\end{enumerate}
In theories with gaugino mass unification and with a lower bound on the
ratio $m_{3/2}/m_{1/2}$, both criteria involve only one mass scale, the
gaugino mass parameter.  Consequently, the upper limit on the NLSP
lifetime can be translated into a lower limit on $m_{1/2}$.  This is a
difference compared to the Constrained MSSM, where $m_{3/2}$ is a free
parameter, so that only the first criterion can be applied
\cite{Pradler:2007is}.

If the lightest neutralino is a pure bino, we can use the approximation
\begin{equation} \label{eq:ApproxMNeutralino}
	m_\chi \approx M_1(M_Z) \approx 
	m_{1/2} \frac{\alpha_1(M_Z)}{\alpha_1(M_\text{GUT})} \approx
	0.42 \, m_{1/2} \;,
\end{equation}
where we have used
$\alpha_1^{-1}(M_Z) \approx 59$ and 
$\alpha_1^{-1}(M_\text{GUT}) \approx 25$.
The approximation for the low-energy value of $M_1$ works very well,
since the running of the gaugino masses is independent of the other soft
parameters at the one-loop level.

We parametrise the minimal gravitino mass as
\begin{equation} \label{eq:m32Andm12}
	m_{3/2}^\text{min} = c \, m_{1/2} \;.
\end{equation}
For example, the mentioned bound from na\"ive dimensional analysis in
gaugino mediation corresponds to $c=0.1$.  If we allow this bound to be
violated by up to an order of magnitude, we obtain the minimal value
$c=0.01$.

Using Eqs.~(\ref{eq:2bodydecay}--\ref{eq:m32Andm12}) with
$m_\slepton=m_\chi$ and $m_{3/2}=m_{3/2}^\text{min}$, we find
\begin{equation}
	\tau_\slepton \approx
 	\frac{48\pi\,c^2\,M_\text{P}^2}{0.42^5\,m_{1/2}^3}
	\left(1-\frac{c^2}{0.42^2}\right)^{-4}
\end{equation}
or, imposing the CBBN bound,
\begin{equation} \label{eq:m12Min}
	m_{1/2} \gtrsim 21\tev \cdot c^{2/3}
	 \left( \frac{\tau_\slepton^\text{max}}{5\cdot10^3\s} \right)^{-\frac{1}{3}}
	 \left( 1+7.6\,c^2 \right) ,
\end{equation}
where we have assumed $c$ to be small.  For instance, $c=0.01$ yields
$m_{1/2} \gtrsim 970\gev$ for $\tau_\slepton^\text{max}=5\cdot10^3\s$.
By setting $m_\slepton=m_\chi$, we have implicitly assumed that this
equality is satisfied in some part of the parameter space.  This is the
case for gravity mediation, NUHM models
\cite{Olechowski:1994gm,Berezinsky:1995cj} or gaugino mediation with
moderate $\tan\beta$ \cite{Buchmuller:2005ma}.  If for a given $m_{1/2}$
the maximal slepton
mass is smaller than the lightest neutralino mass, \Eqref{eq:m12Min}
still holds, but an even stronger limit on $m_{1/2}$ exists.  The same
is true if the lightest neutralino is not a pure bino, since then
$m_\chi$ is smaller than $M_1$.

\subsection{Impact on Gaugino Mediation}
\subsubsection{Numerical Results}
 
Let us now return to the specific setup of gaugino mediation and perform
numerical studies of the impact of CBBN and other cosmological
constraints on the allowed parameter space.  In addition to the upper
limit \eqref{eq:CBBNBound} on the NLSP lifetime, we have to take into
account the non-thermal gravitino abundance resulting from NLSP decays,
\begin{equation} \label{eq:Omeganonth}
	\Omega^\text{non-th}_{3/2}h^2 =
	\frac{m_{3/2}}{m_\text{NLSP}} \Omega_\text{NLSP}^\text{th}h^2 \;.
\end{equation}
For large $m_{1/2}$, it exceeds the observed cold dark matter density,
resulting in an upper limit on $m_{1/2}$.  We use the $95\%$ C.L.\ bound
given in \cite{Hamann:2006pf},
\begin{equation}
	\Omega_\text{DM} h^2 < 0.136 \;.
\end{equation}
Furthermore, there are the ``usual'' BBN constraints on the energy
release from NLSP decays.
With the short lifetime, the electromagnetic energy release is
harmless, but the hadronic energy release becomes relevant with increasing stau
mass.
The calculation for the hadronic branching ratio of right-handed sleptons 
can be found in \cite{Steffen:2006hw}.
We use the hadronic constraints from Fig.~10 of \cite{Jedamzik:2006xz}.
These constraints assume that the whole rest energy of the decaying 
particle ends up in the hadronic shower, which is not the case here.
Rather, the average invariant mass of the $\bar{q}q$ pair emitted in the
hadronic decay $\slepton \rightarrow \ell \gravitino \bar{q} q$
is close to the $Z$ mass, almost independently of the slepton mass \cite{Steffen:2006hw}.
We therefore use the BBN bounds for a decaying particle
of $100\gev$ also for larger NLSP masses and rescale the bound on
$\Omega_\text{NLSP}$ by a
factor $m_\slepton/100\gev$ \cite{Jedamzik}.

Both for the constraints from BBN and for those from the observed
cold dark matter density, the thermal relic density of the NLSP is essential.
We use micrOMEGAs 1.3.7 \cite{Belanger:2001fz,Belanger:2004yn} to
calculate it
numerically.  The superpartner spectrum is determined by 
SOFTSUSY 2.0.14 \cite{Allanach:2001kg}.
For the top quark pole mass, we use
$170.9\gev$ \cite{Tevatron:2007bxa}.\footnote{In addition, we use
$m_b(m_b)=4.25\gev$ and 
$\alpha_s^{\text{SM }\overline{\text{MS}}}(M_Z)=0.1176$ \cite{Yao:2006px}.  
Some other SM parameters are hard-coded in micrOMEGAs, 
$\alpha_\text{em}^{-1 \text{ SM }\overline{\text{MS}}}(M_Z)=127.90896$,
$G_F = 1.16637 \cdot 10^{-5} \gev^{-2}$, and
$m_\tau = 1.777\gev$.}
We restrict ourselves to the case $\tan\beta=10$ and $\mu>0$.

\begin{figure}
  \centering
  \begin{minipage}[b]{7.8 cm}
    \includegraphics[width=7.8cm,height=7cm]{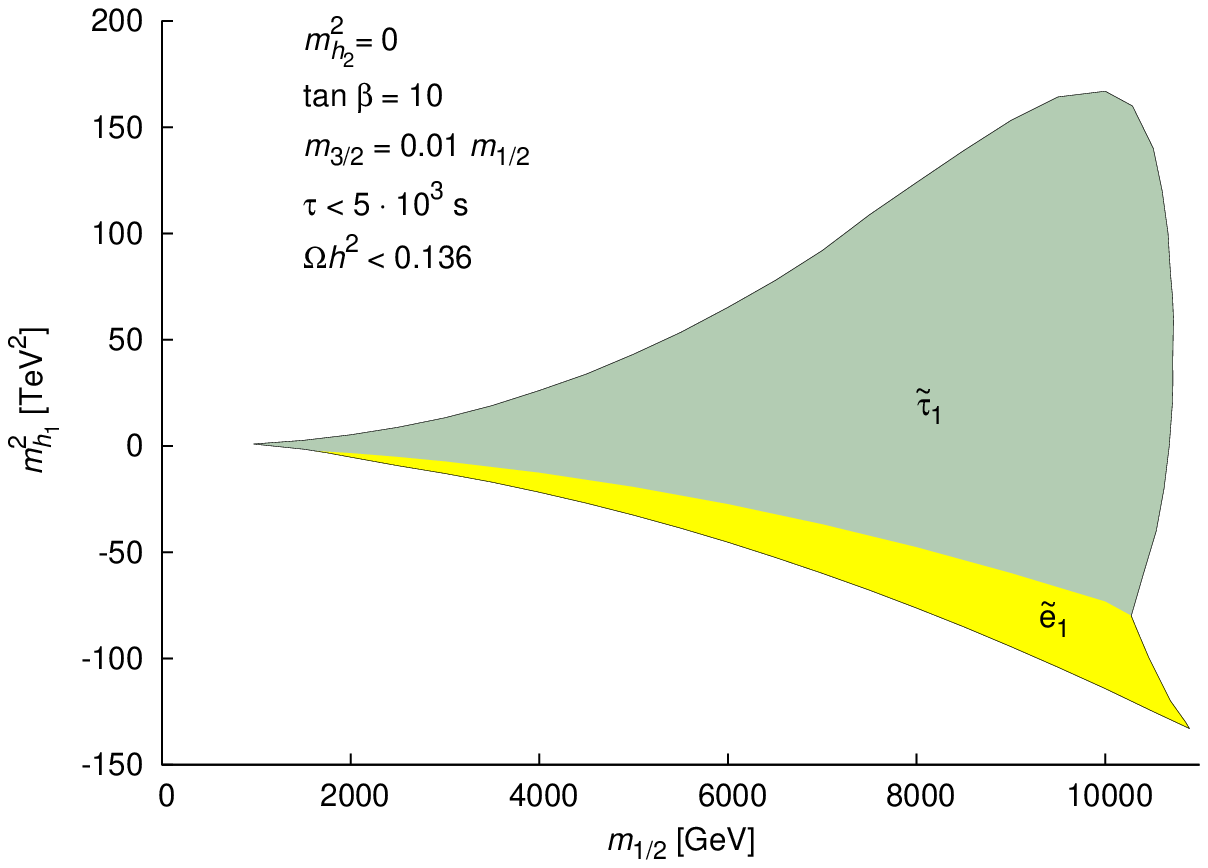}  
  \end{minipage}
  \begin{minipage}[b]{7.8 cm}
    \includegraphics[width=7.8cm,height=7cm]{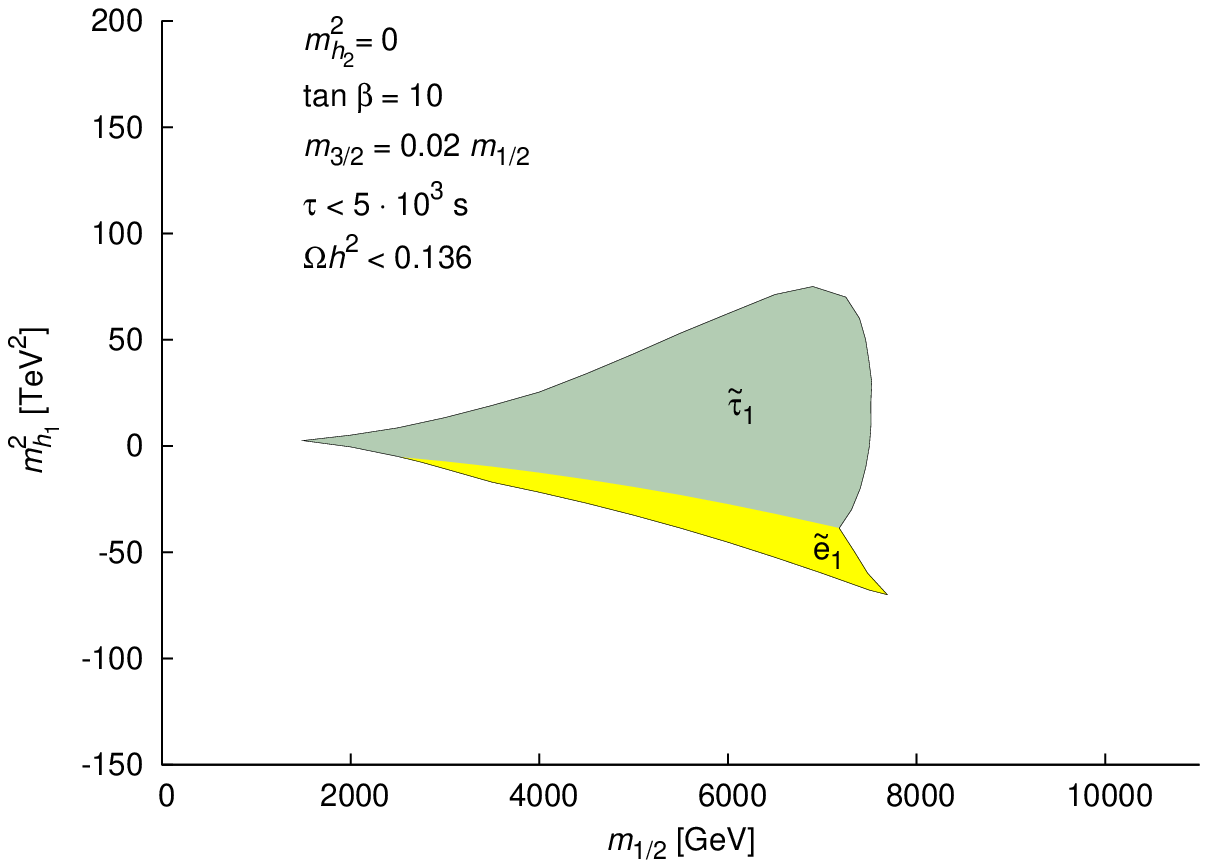}  
  \end{minipage}
  \begin{minipage}[b]{7.8 cm}
    \includegraphics[width=7.8cm,height=7cm]{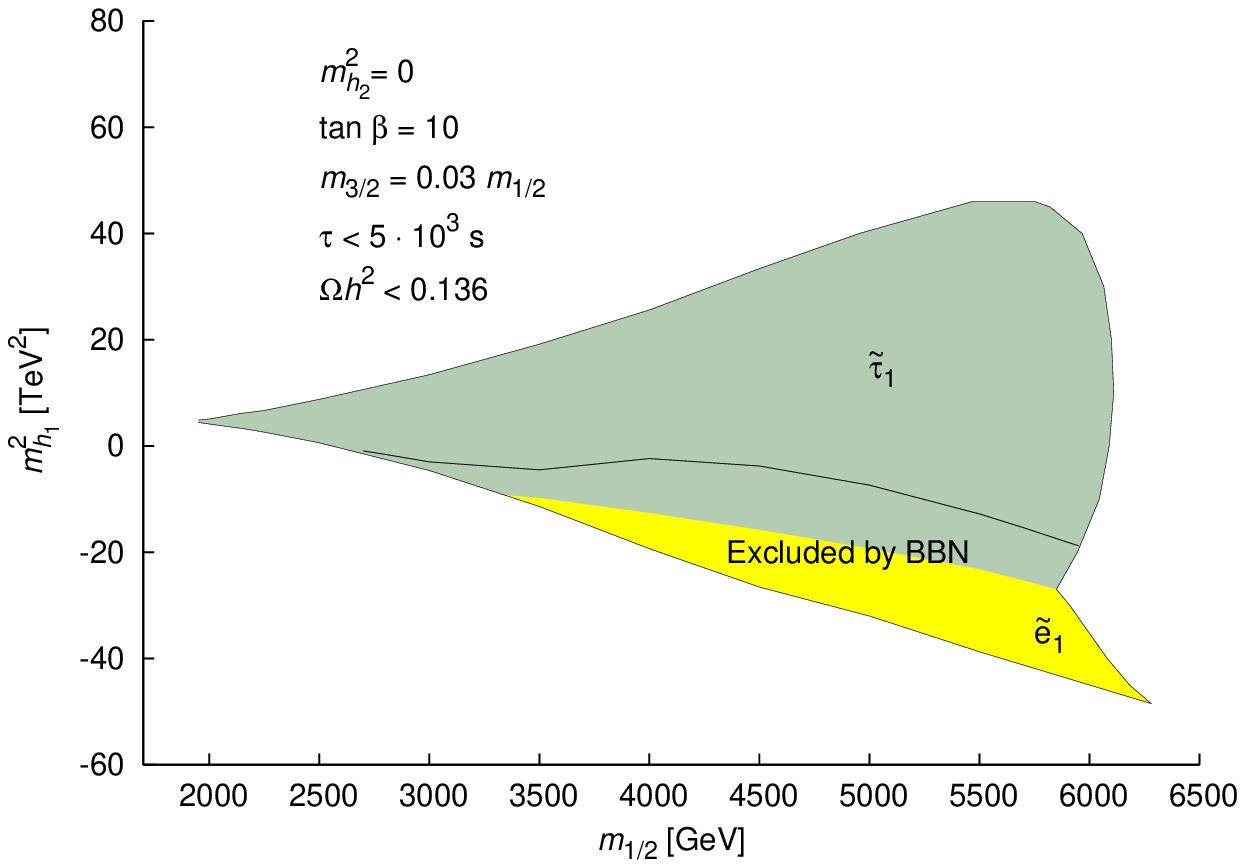}  
  \end{minipage}
  \begin{minipage}[b]{7.8 cm}
    \includegraphics[width=7.8cm,height=7cm]{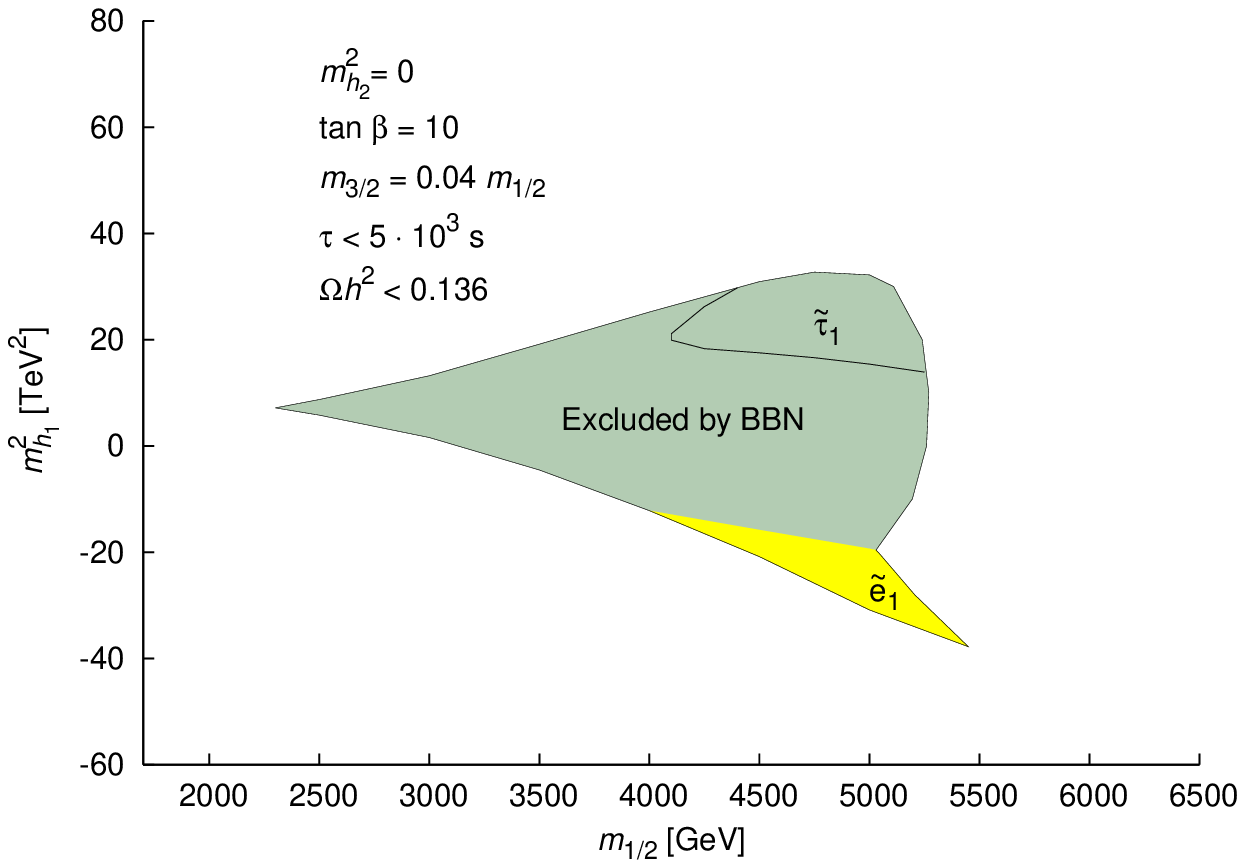}  
  \end{minipage}
  \caption{Allowed regions for $m^2_{h_1}$ and $m_{1/2}$ with
           $m^2_{h_2}=0$,
           $\tan\beta=10$ and different values of the gravitino mass $m_{3/2}$.
           The NLSP is the $\stau_1\approx\stau_\text{R}$ in the green
           (dark-grey) region and the $\widetilde e_\text{R}$ in
           the yellow (light-grey) region.
           The black lines indicate the BBN constraints from hadronic energy release.
           Note that the plots for $c=0.03$ and $c=0.04$ are scaled
           up in comparison to $c=0.01$ and $c=0.02$.
}
  \label{fig:m2h2=0}
\end{figure}

The parameter space in the $m_{1/2}$-$m_{h_1}^2$ plane resulting from the lifetime and cold dark matter 
constraint in addition to constraints from consistency (e.g.\ absence of tachyons)
is shown in \Figref{fig:m2h2=0} for $m^2_{h_2}=0$ and different values of the 
gravitino mass.  The green (dark-grey) area corresponds to a region with
a stau NLSP, while the yellow (light-grey) area corresponds to a region
with a selectron NLSP\@.
The constraint from the observed cold dark matter density restricts the parameter 
space towards large values of $m_{1/2}$, while the lifetime constraint restricts
it towards small $m_{1/2}$.

While for $c=0.01$ and $c=0.02$ only the constraints from overclosure and NLSP lifetime
are relevant, the hadronic BBN constraints become important for $c=0.03$ and $c=0.04$.
For $c \gtrsim 0.05$ there are no allowed regions left.
The excluded region is indicated by the black line.
In the selectron NLSP region, the smuon is only slightly heavier.
Thus, it may directly decay into gravitinos and affect BBN\@.
Consequently, the impact of the BBN constraints on this region may have
been somewhat underestimated in the plots.
For $c=0.01$ we have allowed regions with large $m_{1/2}$ and therefore stau lifetimes
around $10\s$ or less. Here the mesons from $\tau$ decays (where the
$\tau$ stems from the dominant two-body decay $\stau \to \tau\gravitino$)
become relevant, so that the hadronic stau branching ratio is
$\mathcal{O}(1)$. However, one cannot apply the corresponding bound on
$\Omega_\text{NLSP}$ directly here, because it is sensitive to
the number of charged mesons emitted in an NLSP decay \cite{Jedamzik}.
In most tau decay modes there is only one, while a $\bar{q}q$ pair
of $1\tev$, which is assumed in Fig.~9 of \cite{Jedamzik:2006xz}, results in around $25$ mesons \cite{Jedamzik}.
Therefore, we relaxed the bound from this figure by
a factor $25 \, m_\stau/1\tev$. 
Applying the resulting limit puts no additional constraints on the parameter space
of gaugino mediation.
Also the cosmic microwave background 
does not yield constraints for
$\tau_\stau < 10^5\s$ \cite{Lamon:2005jc}.

Let us now turn to non-zero values of $m_{h_2}^2$.  While increasing
$m_{1/2}$, we also increase this soft mass in such a way that the ratio
$m_{h_2}^2/m_{1/2}^2$ remains fixed.  Otherwise, any value of
$m_{h_2}^2$ allowed for smaller $m_{1/2}$ would become completely
irrelevant at large $m_{1/2}$.  As mentioned earlier, the slepton masses
are determined mainly by $m^2_{h_1}-m^2_{h_2}$, and the lightest
neutralino (bino) mass is almost independent of the soft Higgs masses if
$m_{h_2}^2$ is not too large.  Consequently, the effect of a non-zero
but moderate value of $|m_{h_2}^2|/m_{1/2}^2$ is simply a vertical shift
(to larger or smaller values of $m_{h_1}^2$) of the allowed parameter
space region, and therefore we do not show any examples.

\begin{figure}
  \centering
  \begin{minipage}[b]{7.8 cm}
    \includegraphics[width=7.8cm,height=7cm]{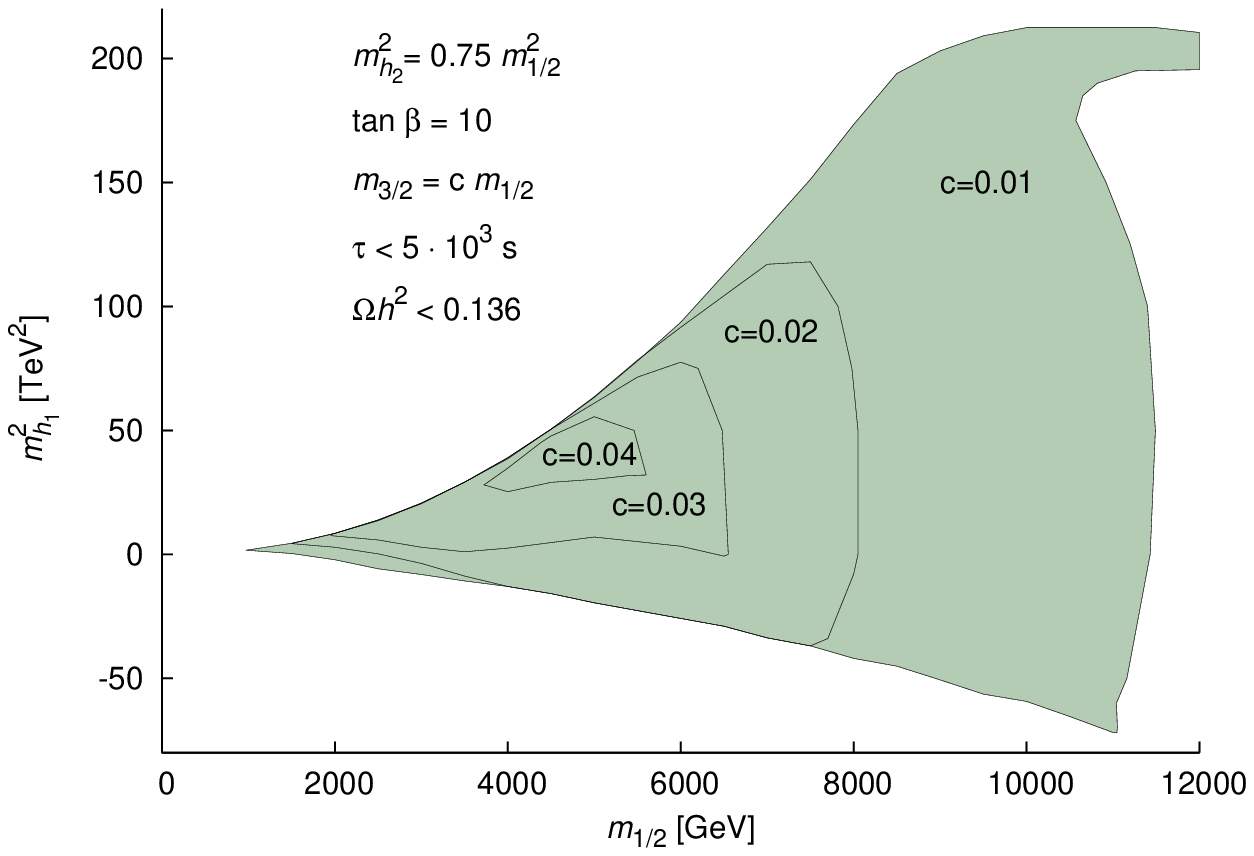}  
  \end{minipage}
  \begin{minipage}[b]{7.8 cm}
    \includegraphics[width=7.8cm,height=7cm]{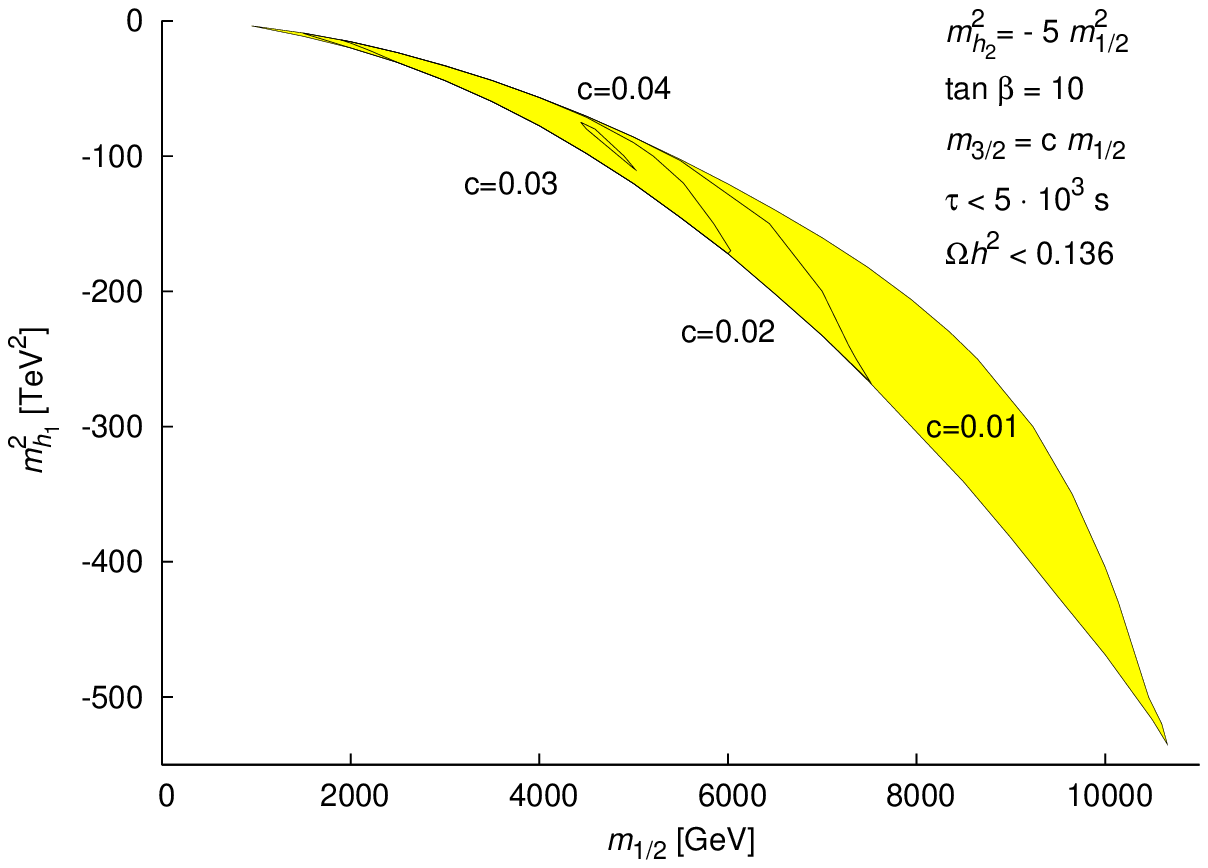}  
  \end{minipage}
  \caption{Allowed regions for $m^2_{h_1}$ and $m_{1/2}$ with
$m^2_{h_2}=0.75 \, m^2_{1/2}$ (left)
           and $m^2_{h_2}=-5 \,m^2_{1/2}$ (right) for different values of the parameter $c$. 
           In the left panel we have a stau NLSP only,
           while in the right panel the NLSP is a selectron.
           The black lines indicate the allowed regions for different values of $c$.
}
  \label{fig:m2h2neq0}
\end{figure}

For rather large positive $m_{h_2}^2$, the lightest
neutralino becomes lighter due to a significant higgsino admixture.  For
negative values, there is only little space between the unphysical
region and the neutralino LSP domain.  These effects are illustrated
in \Figref{fig:m2h2neq0}.
For $m_{h_2}^2=0.75\,m_{1/2}^2$, which is the maximal value allowed for
all values of $m_{1/2}$ we consider, we always find a stau NLSP\@.
The ``trunk'' in the  upper right corner of the allowed region for
this case is due to coannihilations with higgsinos,
which reduce the stau abundance.  
For $m_{h_2}^2=-5\,m_{1/2}^2$, close to the limit on the other side
of the parameter space, we have a selectron NLSP\@. 
As in the case of $m^2_{h_2}=0$, only the constraints from overclosure and lifetime
are relevant for $c=0.01$ and $c=0.02$, while for $c=0.03$ and $c=0.04$ the hadronic
BBN constraints become important. Again, for $c=0.05$ there is no valid region left.

In summary, we conclude that catalyzed primordial nucleosynthesis as well as other
cosmological constraints place an upper
bound on the gravitino mass in the $\slepton$ region of gaugino
mediation,
$m_{3/2} < 0.05\, m_{1/2}$.

\subsubsection{Consequences for the Superparticle Mass Spectrum}
\Tabref{tab:m12Min} shows an overview of the minimal values we find for
$m_{1/2}$.
\begin{table}
\centering
\begin{tabular}{|c|cccc|}
\hline
$c$ & 0.01 & 0.02 & 0.03 & 0.04 \\
\hline
\Eqref{eq:m12Min} & 970 & 1540 & 2020 & 2460 \\
$\tau_\slepton^\text{max}=5\cdot10^3\s$ & 960 & 1480 & 1910 & 4100 (BBN) \\
$\tau_\slepton^\text{max}=1\cdot10^4\s$ & 780 & 1190 & 1540 & \\
$\tau_\slepton^\text{max}=2\cdot10^3\s$ & 1270 & 1970 & 2540 & \\
$\tau_\slepton^\text{max}=1\cdot10^3\s$ & 1580 & 2440 & 3150 & \\
\hline
\end{tabular}
\caption{Lower limits on $m_{1/2}$ in \Gev.  The numerical bound for
 $c=0.04$ stems from the BBN constraint on energy release from NLSP
 decays, while the remaining bounds are due to the limit
 $m_\slepton<m_\slepton^\text{max}$ from CBBN\@.  The values in the second
 line were obtained from the analytical estimate \eqref{eq:m12Min} for
 $\tau_\slepton^\text{max}=5\cdot10^3\s$.  We set $m^2_{h_2}=0$ and 
 $\tan\beta=10$ in all cases.
}
\label{tab:m12Min}
\end{table}
We see that \Eqref{eq:m12Min} works with an accuracy of a few percent.
It can be further improved by evaluating $\alpha_1$ at the bino mass,
i.e.\ close to a \Tev{} for larger values of $m_{1/2}$.  The
results depend only weakly on $m_{h_2}^2$, varying by not more
than $2\%$ as long as $m_{h_2}^2$ does not lie very close to the border
of the allowed parameter space.  A moderate increase of $\tan\beta$ 
by a factor $\sim 2$ does
not have a big effect either, since the decrease of the stau mass due to
the larger Yukawa coupling can be compensated by raising $m_{h_1}^2$,
and analogously for a decreased $\tan\beta$.  However, for
$\tan\beta\gtrsim30$, a
charged slepton is always lighter than the lightest neutralino for
$m_{h_2}^2\sim0$ \cite{Buchmuller:2005ma,Evans:2006sj}, so that larger
values of $m_{1/2}$ are required.  To give an impression of the effect
that a change of the lifetime constraint from CBBN can have, we also
show some examples with different values of the maximal slepton
lifetime in the table.

At the points in parameter space where $m_{1/2}$ takes its minimal value
for different values of $c$, we obtain the mass spectra given in
\Tabref{tab:Spectra}.
\begin{table}
\centering
\begin{tabular}{|l|ccc|}
\hline
$c$											   & $0.01$	& $0.02$	& $0.03$\\
\hline
$m_{1/2}^\text{min}$ & 960 & 1480 & 1910 \\
$m_{h_1}^2/\tev^2$ & 0.88 & 2.53 & 4.64 \\
\hline
$\widetilde g$                                 & 2096		& 3130		& 3972\\
Other $\widetilde q$                           & 1755\,--\,1902 & 2613\,--\,2827& 3311\,--\,3578\\
$\widetilde t_1$                               & 1485 		& 2217		& 2808\\
$\chi^\pm_2, \chi^0_3, \chi^0_4$               & 1107\,--\,1112	& 1605\,--\,1612& 2002\,--\,2012\\
$\chi^\pm_1$                                   & 769 		& 1198		& 1555\\
$\chi^0_2$                                     & 763 		& 1188		& 1543\\
$\widetilde e_\text{L}, \widetilde\mu_\text{L}$& 623 		& 943		& 1205\\
$\stau_2$                                      & 620 		& 937		& 1197\\
$\snu_e, \snu_\mu, \snu_\tau$                  & 610\,--\,614	& 927\,--\,934	& 1187\,--\,1196\\
$\widetilde e_\text{R}, \widetilde\mu_\text{R}$& 418 		& 655		& 855\\
$\chi^0_1$                                     & 405 		& 635		& 830\\
$\stau_1$                                      & 405		& 635		& 829\\
$\gravitino$								   & 9.6 		& 29.5		& 57.2\\
\hline
\end{tabular}
\caption{Superparticle mass spectra corresponding to the minimal
 $m_{1/2}$ allowed by the CBBN
 constraint~\eqref{eq:CBBNBound} for $\tan\beta=10$ and $m_{h_2}^2=0$.
 All masses are given in \Gev{} unless stated otherwise.  ``Other
 $\widetilde q\,$'' refers to all squarks other than $\widetilde t_1$.
}
\label{tab:Spectra}
\end{table}
Again, there is little variation for different values of $m_{h_2}^2$, as
long as they are not too close to the border of the allowed region.
According to \cite{Kazana:1999,Acosta:2002wf}, LHC will be able to find
long-lived staus with masses up to around $700\gev$.  Thus, for 
$m_{3/2} \lesssim 0.02\,m_{1/2}$, it should be possible to detect
supersymmetry at least in a part of the allowed slepton NLSP region.
Compared to similar points in the Constrained MSSM, the slepton spectrum
is compressed (i.e.\ the difference
between the masses of $\slepton_\text{L}$ and $\slepton_\text{R}$ is
smaller) due to the non-zero $m_{h_1}^2$.

\subsubsection{Constraints on the Reheating Temperature}

At high temperatures, gravitinos are produced by thermal scatterings.  The
resulting energy density is approximately given by 
\cite{Bolz:2000fu,Pradler:2006qh}
\begin{equation}
	\Omega_{3/2}^\text{th} h^2 \simeq
	0.27 \left( \frac{T_\text{R}}{10^{10}\gev} \right)
	\left( \frac{100\gev}{m_{3/2}} \right)
	\biggl( \frac{m_{\tilde g}}{1\tev} \biggr)^2 \;,
\end{equation}
where $m_{\tilde g}$ is the running gluino mass evaluated at low energy.

We can obtain a constraint on the reheating temperature 
using $m_{3/2}=c\cdot m_{1/2}$, $m_{\tilde{g}} \sim
\tfrac{\alpha_s(M_Z)}{\alpha_s(M_\text{GUT})}\, m_{1/2} \sim 2.9 \,m_{1/2}$ and
$\Omega_\text{DM} h^2 < 0.136$,
\begin{equation} \label{TR}
T_\text{R} \lesssim 5.8 \, c \left(\frac{\tev}{m_{1/2}}\right) 10^9 \gev \;.
\end{equation}
The corresponding maximal reheating temperatures for our values of $c$ and $m_{1/2}$
are given in Tab.~\ref{tab:TR}.
The results are similar to those obtained in the Constrained MSSM for
gravitino masses of a similar order of magnitude \cite{Pradler:2007is}.
Note that we have not taken into account the non-thermally produced
gravitino density \eqref{eq:Omeganonth} here.
These values imply \cite{Buchmuller:2004nz} that generically thermal
leptogenesis is not possible in gaugino mediation with charged sleptons
as NLSPs%
\footnote{See, however, \cite{Raidal:2004vt} for a special setup where
thermal leptogenesis also works for low $T_\text{R}$.
},
unless there is entropy production between the stau decoupling and primordial
nucleosynthesis. 
\begin{table}
\centering 
\renewcommand{\arraystretch}{1.1}
\begin{tabular}{|c|cccc|}
\hline
$c$ & 0.01 & 0.02 & 0.03 & 0.04 \\
\hline
$m^\text{min}_{1/2}/\Gev$ & 960 & 1480 & 1910 & 4100 \\
\hline
$T^\text{max}_\text{R}/\Gev$ & $6\cdot 10^7$ & $8\cdot 10^7$ & $9\cdot 10^7$ & $6\cdot 10^7$ \\
\hline
\end{tabular}
\caption{Upper limits on the reheating temperature $T_\text{R}$ in gaugino mediation
with a stau NLSP.   
}
\label{tab:TR}
\end{table}

\subsubsection{Left-Handed Stau NLSPs}
For larger values of $\tan\beta$, there is a parameter space region where a
predominantly ``left-handed'' stau is the NLSP\@.
Since the decay rate is the same for left- and right-handed staus,
the CBBN constraint resulting from the lifetime of the NLSP is similar. 
However, there will be a difference in the constraints from hadronic decays, since
the hadronic branching ratio is considerably larger for left-handed 
staus. Unfortunately no detailed calculation for this branching ratio
has been performed so far, but the result should be similar to the case of left-handed 
sneutrinos \cite{Kanzaki:2006hm}.
Despite the larger hadronic branching ratio a rough estimation indicates that
there will be allowed regions also in the case of a left-handed slepton
NLSP\@.
We leave the detailed discussion of this region for future work.

\section{Conclusions}

We have discussed cosmological constraints on theories with a gravitino
LSP and a charged slepton NLSP\@.  In particular, the recently
discovered effect of Catalyzed BBN places a stringent upper limit on the
NLSP lifetime.  From this, we have derived a lower limit on the unified
gaugino mass parameter $m_{1/2}$ for scenarios with a lower bound
$m_{3/2} > c \, m_{1/2}$ on the gravitino mass.  We have numerically
determined the part of the parameter space of gaugino mediation with a
charged slepton NLSP that remains compatible with all constraints from
BBN and the observed dark matter density.  Allowed regions exist for
$c<0.05$, which means that the gravitino mass bound from
na\"ive dimensional analysis, corresponding to $c\sim0.1$, has to be
violated by a factor of at least 2 to 3.  If we set a conservative lower
limit of $c\gtrsim0.01$, $m_{1/2}$ may be as small as $1\tev$, so that
supersymmetry can still be within the discovery reach of the LHC\@.

Smaller superparticle masses can be viable, if one relaxes the
assumptions on the cosmological scenario.  For example, entropy
production between NLSP freeze-out and the start of BBN can dilute the
NLSP abundance sufficiently to satisfy all constraints even for long
lifetimes
\cite{Buchmuller:2006tt,Pradler:2006hh,Hamaguchi:2007mp,Kasuya:2007cy}.
Alternatively, a
reheating temperature significantly below the NLSP mass can result in a
suppressed NLSP abundance, too \cite{Takayama:2007du}.

\section*{Acknowledgements}
We would like to thank Bobby Acharya, Ben Allanach, Wilfried
Buchm\"uller, Michael Ratz, the micrOMEGAs team, and Piero Ullio
for valuable discussions.
Special thanks go to Karsten Jedamzik for explaining the application
of the BBN constraints resulting from hadronic NLSP decays.
This work has been supported by the SFB-Transregio 27 ``Neutrinos and 
Beyond'' and by the DFG cluster of excellence ``Origin and 
Structure of the Universe''.
KSH would like to thank the Abdus Salam ICTP for hospitality during
a part of this work.

\bibliography{GauginoMediation}

\providecommand{\bysame}{\leavevmode\hbox to3em{\hrulefill}\thinspace}
\frenchspacing
\newcommand{\origttfamily}{}
\let\origttfamily=\ttfamily
\renewcommand{\ttfamily}{\origttfamily \hyphenchar\font=`\-}

\begin{thebibliography}{10}

\bibitem{Falomkin:1984eu}
I.~V. Falomkin et~al., Nuovo Cim. \textbf{A79} (1984), 193, [Yad.\ Fiz.\ {\bf
  39} (1984), 990].

\bibitem{Khlopov:1984pf}
M.~Y. Khlopov and A.~D. Linde, Phys. Lett. \textbf{B138} (1984), 265.

\bibitem{Ellis:1984eq}
J.~R. Ellis, J.~E. Kim, and D.~V. Nanopoulos, Phys. Lett. \textbf{B145} (1984),
  181.

\bibitem{Kawasaki:2004qu}
M.~Kawasaki, K.~Kohri, and T.~Moroi, Phys. Rev. \textbf{D71} (2005), 083502,
  \texttt{astro-ph/0408426}.

\bibitem{Feng:2004mt}
J.~L. Feng, S.~Su, and F.~Takayama, Phys. Rev. \textbf{D70} (2004), 075019,
  \texttt{hep-ph/0404231}.

\bibitem{Cerdeno:2005eu}
D.~G. Cerde\~{n}o, K.-Y. Choi, K.~Jedamzik, L.~Roszkowski, and R.~Ruiz~de
  Austri, JCAP \textbf{0606} (2006), 005, \texttt{hep-ph/0509275}.

\bibitem{Kanzaki:2006hm}
T.~Kanzaki, M.~Kawasaki, K.~Kohri, and T.~Moroi, Phys. Rev. \textbf{D75}
  (2007), 025011, \texttt{hep-ph/0609246}.

\bibitem{Covi:2007xj}
L.~Covi and S.~Kraml, JHEP \textbf{08} (2007), 015, \texttt{hep-ph/0703130}.

\bibitem{Buchmuller:2004rq}
W.~Buchm{\"u}ller, K.~Hamaguchi, M.~Ratz, and T.~Yanagida, Phys. Lett.
  \textbf{B588} (2004), 90, \texttt{hep-ph/0402179}.

\bibitem{Albuquerque:2003mi}
I.~Albuquerque, G.~Burdman, and Z.~Chacko, Phys.\ Rev.\ Lett. \textbf{92}
  (2004), 221802, \texttt{hep-ph/0312197}.

\bibitem{Fujii:2003nr}
M.~Fujii, M.~Ibe, and T.~Yanagida, Phys. Lett. \textbf{B579} (2004), 6,
  \texttt{hep-ph/0310142}.

\bibitem{Ellis:2003dn}
J.~R. Ellis, K.~A. Olive, Y.~Santoso, and V.~C. Spanos, Phys. Lett.
  \textbf{B588} (2004), 7, \texttt{hep-ph/0312262}.

\bibitem{Roszkowski:2004jd}
L.~Roszkowski, R.~Ruiz~de Austri, and K.-Y. Choi, JHEP \textbf{08} (2005), 080,
  \texttt{hep-ph/0408227}.

\bibitem{Jedamzik:2005dh}
K.~Jedamzik, K.-Y. Choi, L.~Roszkowski, and R.~Ruiz~de Austri, JCAP
  \textbf{0607} (2006), 007, \texttt{hep-ph/0512044}.

\bibitem{Steffen:2006hw}
F.~D. Steffen, JCAP \textbf{09} (2006), 001, \texttt{hep-ph/0605306}.

\bibitem{Buchmuller:2006nx}
W.~Buchm{\"u}ller, L.~Covi, J.~Kersten, and K.~Schmidt-Hoberg, JCAP
  \textbf{0611} (2006), 007, \texttt{hep-ph/0609142}.

\bibitem{Cyburt:2006uv}
R.~H. Cyburt, J.~Ellis, B.~D. Fields, K.~A. Olive, and V.~C. Spanos, JCAP
  \textbf{0611} (2006), 014, \texttt{astro-ph/0608562}.

\bibitem{Pradler:2006hh}
J.~Pradler and F.~D. Steffen, Phys. Lett. \textbf{B648} (2007), 224,
  \texttt{hep-ph/0612291}.

\bibitem{Pospelov:2006sc}
M.~Pospelov, Phys. Rev. Lett. \textbf{98} (2007), 231301,
  \texttt{hep-ph/0605215}.

\bibitem{Kohri:2006cn}
K.~Kohri and F.~Takayama, Phys. Rev. \textbf{D76} (2007), 063507,
  \texttt{hep-ph/0605243}.

\bibitem{Kaplinghat:2006qr}
M.~Kaplinghat and A.~Rajaraman, Phys. Rev. \textbf{D74} (2006), 103004,
  \texttt{astro-ph/0606209}.

\bibitem{Hamaguchi:2007mp}
K.~Hamaguchi, T.~Hatsuda, M.~Kamimura, Y.~Kino, and T.~T. Yanagida, Phys. Lett.
  \textbf{B650} (2007), 268, \texttt{hep-ph/0702274}.

\bibitem{Bird:2007ge}
C.~Bird, K.~Koopmans, and M.~Pospelov, \texttt{hep-ph/0703096}.

\bibitem{Kawasaki:2007xb}
M.~Kawasaki, K.~Kohri, and T.~Moroi, Phys. Lett. \textbf{B649} (2007), 436,
  \texttt{hep-ph/0703122}.

\bibitem{Jedamzik:2007cp}
K.~Jedamzik, \texttt{arXiv:0707.2070 [astro-ph]}.

\bibitem{Pradler:2007is}
J.~Pradler and F.~D. Steffen, \texttt{arXiv:0710.2213 [hep-ph]}.

\bibitem{Kaplan:1999ac}
D.~E. Kaplan, G.~D. Kribs, and M.~Schmaltz, Phys. Rev. \textbf{D62} (2000),
  035010, \texttt{hep-ph/9911293}.

\bibitem{Chacko:1999mi}
Z.~Chacko, M.~A. Luty, A.~E. Nelson, and E.~Ponton, JHEP \textbf{01} (2000),
  003, \texttt{hep-ph/9911323}.

\bibitem{Buchmuller:2005ma}
W.~Buchm{\"u}ller, J.~Kersten, and K.~Schmidt-Hoberg, JHEP \textbf{02} (2006),
  069, \texttt{hep-ph/0512152}.

\bibitem{Evans:2006sj}
J.~L. Evans, D.~E. Morrissey, and J.~D. Wells, Phys. Rev. \textbf{D75} (2007),
  055017, \texttt{hep-ph/0611185}.

\bibitem{Ellis:2002iu}
J.~R. Ellis, T.~Falk, K.~A. Olive, and Y.~Santoso, Nucl. Phys. \textbf{B652}
  (2003), 259, \texttt{hep-ph/0210205}.

\bibitem{Baer:2004fu}
H.~Baer, A.~Mustafayev, S.~Profumo, A.~Belyaev, and X.~Tata, Phys. Rev.
  \textbf{D71} (2005), 095008, \texttt{hep-ph/0412059}.

\bibitem{Falk:1995cq}
T.~Falk, K.~A. Olive, L.~Roszkowski, and M.~Srednicki, Phys. Lett.
  \textbf{B367} (1996), 183, \texttt{hep-ph/9510308}.

\bibitem{Riotto:1995am}
A.~Riotto and E.~Roulet, Phys. Lett. \textbf{B377} (1996), 60,
  \texttt{hep-ph/9512401}.

\bibitem{Kusenko:1996jn}
A.~Kusenko, P.~Langacker, and G.~Segre, Phys. Rev. \textbf{D54} (1996), 5824,
  \texttt{hep-ph/9602414}.

\bibitem{Falk:1996zt}
T.~Falk, K.~A. Olive, L.~Roszkowski, A.~Singh, and M.~Srednicki, Phys. Lett.
  \textbf{B396} (1997), 50, \texttt{hep-ph/9611325}.

\bibitem{Chacko:1999hg}
Z.~Chacko, M.~A. Luty, and E.~Ponton, JHEP \textbf{07} (2000), 036,
  \texttt{hep-ph/9909248}.

\bibitem{Buchmuller:2005rt}
W.~Buchm{\"u}ller, K.~Hamaguchi, and J.~Kersten, Phys. Lett. \textbf{B632}
  (2006), 366, \texttt{hep-ph/0506105}.

\bibitem{Hebecker:2004ce}
A.~Hebecker and M.~Trapletti, Nucl. Phys. \textbf{B713} (2005), 173,
  \texttt{hep-th/0411131}.

\bibitem{Schmaltz:2000gy}
M.~Schmaltz and W.~Skiba, Phys. Rev. \textbf{D62} (2000), 095005,
  \texttt{hep-ph/0001172}.

\bibitem{Schmaltz:2000ei}
M.~Schmaltz and W.~Skiba, Phys. Rev. \textbf{D62} (2000), 095004,
  \texttt{hep-ph/0004210}.

\bibitem{Olechowski:1994gm}
M.~Olechowski and S.~Pokorski, Phys. Lett. \textbf{B344} (1995), 201,
  \texttt{hep-ph/9407404}.

\bibitem{Berezinsky:1995cj}
V.~Berezinsky et~al., Astropart. Phys. \textbf{5} (1996), 1,
  \texttt{hep-ph/9508249}.

\bibitem{Hamann:2006pf}
J.~Hamann, S.~Hannestad, M.~S. Sloth, and Y.~Y.~Y. Wong, Phys. Rev.
  \textbf{D75} (2007), 023522, \texttt{astro-ph/0611582}.

\bibitem{Jedamzik:2006xz}
K.~Jedamzik, Phys. Rev. \textbf{D74} (2006), 103509, \texttt{hep-ph/0604251}.

\bibitem{Jedamzik}
K.~Jedamzik, private communication.

\bibitem{Belanger:2001fz}
G.~Belanger, F.~Boudjema, A.~Pukhov, and A.~Semenov, Comput. Phys. Commun.
  \textbf{149} (2002), 103, \texttt{hep-ph/0112278}.

\bibitem{Belanger:2004yn}
G.~Belanger, F.~Boudjema, A.~Pukhov, and A.~Semenov, Comput. Phys. Commun.
  \textbf{174} (2006), 577, \texttt{hep-ph/0405253}.

\bibitem{Allanach:2001kg}
B.~C. Allanach, Comput. Phys. Commun. \textbf{143} (2002), 305,
  \texttt{hep-ph/0104145}.

\bibitem{Tevatron:2007bxa}
CDF and D0 Collaborations, {Tevatron Electroweak Working Group},
  \texttt{hep-ex/ 0703034}.

\bibitem{Yao:2006px}
Particle Data Group, W.~M. Yao et~al., J. Phys. \textbf{G33} (2006), 1.

\bibitem{Lamon:2005jc}
R.~Lamon and R.~Durrer, Phys. Rev. \textbf{D73} (2006), 023507,
  \texttt{hep-ph/0506229}.

\bibitem{Kazana:1999}
M.~Kazana, G.~Wrochna, and P.~Zalewski, CMS CR 1999/019,\\
  \texttt{http://cms.cern.ch/iCMS/}.

\bibitem{Acosta:2002wf}
ATLAS and CMS Collaborations, D.~Acosta, in \emph{Proceedings
  of the 14th Topical Conference on Hadron Collider Physics (HCP 2002)}
  (M.~Erdmann and T.~M{\"u}ller, eds.), Springer, Berlin (2003),\\
  \texttt{http://hcp2002.physik.uni-karlsruhe.de/talks/Th\_15\_Acosta.pdf}.

\bibitem{Bolz:2000fu}
M.~Bolz, A.~Brandenburg, and W.~Buchm{\"u}ller, Nucl. Phys. \textbf{B606}
  (2001), 518, \texttt{hep-ph/0012052}.

\bibitem{Pradler:2006qh}
J.~Pradler and F.~D. Steffen, Phys. Rev. \textbf{D75} (2007), 023509,
  \texttt{hep-ph/0608344}.

\bibitem{Buchmuller:2004nz}
W.~Buchm{\"u}ller, P.~Di~Bari, and M.~Pl{\"u}macher, Ann. Phys. \textbf{315}
  (2005), 305, \texttt{hep-ph/0401240}.

\bibitem{Raidal:2004vt}
M.~Raidal, A.~Strumia, and K.~Turzy{\'n}ski, Phys. Lett. \textbf{B609}
(2005),
  351 [Erratum ibid.\ {\bf B632} (2006), 752], \texttt{hep-ph/0408015}.

\bibitem{Buchmuller:2006tt}
W.~Buchm{\"u}ller, K.~Hamaguchi, M.~Ibe, and T.~T. Yanagida, Phys. Lett.
  \textbf{B643} (2006), 124, \texttt{hep-ph/0605164}.

\bibitem{Kasuya:2007cy}
S.~Kasuya and F.~Takahashi, \texttt{arXiv:0709.2634 [hep-ph]}.

\bibitem{Takayama:2007du}
F.~Takayama, \texttt{arXiv:0704.2785 [hep-ph]}.

\end{thebibliography}
\bibliographystyle{NewArXiv}

\end{document}